\documentclass[aps,pra,superscriptaddress,twocolumn]{revtex4}

\usepackage{comment}
\usepackage[utf8]{inputenc}
\usepackage{amssymb}
\usepackage{amsmath}
\usepackage{amsfonts}
\usepackage{bm}
\usepackage{mathtools}
\usepackage{bbold}
\usepackage{upgreek}
\usepackage{xfrac}
\usepackage{soul}
\usepackage{bm}
\usepackage{comment}
\usepackage{xcolor}
\usepackage{mathtools}

\newcommand\widebar[1]{\mathop{\overline{#1}}}

\usepackage[colorlinks=true,citecolor=blue]{hyperref}
\hypersetup{colorlinks=true,citecolor=blue,linkcolor=blue,urlcolor=blue}
\usepackage{url}

\usepackage{txfonts}
\DeclareSymbolFont{matha}{OML}{txmi}{m}{it}
\DeclareMathSymbol{\varv}{\mathord}{matha}{118}
\usepackage{bbm}

\begin{document}

\title{Generalized quantum circuit differentiation rules}

\author{Oleksandr Kyriienko}
\affiliation{Department of Physics and Astronomy, University of Exeter, Stocker Road, Exeter EX4 4QL, United Kingdom}
\affiliation{Qu \& Co B.V., PO Box 75872, 1070 AW, Amsterdam, The Netherlands}

\author{Vincent E. Elfving}
\affiliation{Qu \& Co B.V., PO Box 75872, 1070 AW, Amsterdam, The Netherlands}

\date{\today}

\begin{abstract}
Variational quantum algorithms that are used for quantum machine learning rely on the ability to automatically differentiate parametrized quantum circuits with respect to underlying parameters. Here, we propose the rules for differentiating quantum circuits (unitaries) with arbitrary generators. Unlike the standard parameter shift rule valid for unitaries generated by operators with spectra limited to at most two unique eigenvalues (represented by involutory and idempotent operators), our approach also works for generators with a generic non-degenerate spectrum. Based on a spectral decomposition, we derive a simple recipe that allows explicit derivative evaluation. The derivative corresponds to the weighted sum of measured expectations for circuits with shifted parameters. The number of function evaluations is equal to the number of unique positive non-zero spectral gaps (eigenvalue differences) for the generator. We apply the approach to relevant examples of two-qubit gates, among others showing that the fSim gate can be differentiated using four measurements. Additionally, we present generalized differentiation rules for the case of Pauli string generators, based on distinct shifts (here named as the triangulation approach), and analyze the variance for derivative measurements in different scenarios. Our work offers a toolbox for the efficient hardware-oriented differentiation needed for circuit optimization and operator-based derivative representation.
\end{abstract}

\maketitle

\section{Introduction}
Quantum computing offers a powerful toolbox for solving linear algebra problems~\cite{HHL2009,Childs2017}. It is considered as a prime candidate for improving data analysis and revolutionizing machine learning (ML)~\cite{Lloyd2014,Rebenfrost2014,SchuldRev}. The emergent field of quantum machine learning (QML) gained significant attention in the last several years~\cite{Biamonte2017}. From the application perspective, QML algorithms for classification~\cite{Havlicek2019,SchuldBocharov2020,Du2021,Xia2021,Nghiem2021,Samuel2021}, generative modelling~\cite{JGLiu2018,Zeng2019,Benedetti2019b,Zoufal2019,Coyle2020,Romero2021,Huang2020}, reinforcement learning~\cite{DunjkoRev,Melnikov2018,Saggio2021}, and solving differential equations~\cite{Lubasch2020,Kyriienko2021,Knudsen2020,Garcia-Molina2021} were proposed recently. From the operational perspective, the algorithmic workflow has changed from a fault tolerance-oriented approach with deep ancilla-based circuits~\cite{Biamonte2017} to a hybrid quantum-classical approach~\cite{Benedetti2019rev,CerezoRev}. The former approach guarantees an algorithmic speed-up, but is resource-intensive and demands large scale quantum computers~\cite{Rebenfrost2014}. The latter approach is widely represented by variational quantum algorithms (VQA)~\cite{CerezoRev,Cerezo2021} that may offer advantage for near- and midterm noisy devices~\cite{BhartiRev}.

To perform QML and solve relevant problems, first the data (quantum or classical) shall be encoded in a quantum register~\cite{Benedetti2019rev}. For classical data this can be achieved by using quantum feature maps~\cite{SchuldPRL,Goto2020,Schuld2021}, where variables $x$ are embedded through (nonlinearly transformed) phases of rotations or Hamiltonian evolution, represented by a unitary operator $\hat{U}(x)$. Next, following the VQA workflow, data is manipulated by a trainable parameterized circuit. Much like the model weights in classical neural networks, the parameters $\bm{\theta}$ (often phases of unitary operators that form the circuit, $\hat{V}_{\bm{\theta}} = \prod_{k=1}^{K} \hat{V}_{\theta_k}$) are adjusted variationally based on a training objective defined by a loss function~\cite{Benedetti2019rev,CerezoRev}. Similar to classical machine learning, this is a non-convex optimization problem that requires a stable optimization schedule. In the field of deep learning this was achieved by utilizing stochastic gradient descent and automatic differentiation (AD)~\cite{LeCun2015}. They crucially improve the navigation in a rugged/high-dimensional loss landscape. For QML, given the complexity of quantum optimization for data embedded in a Hilbert space and challenges of noisy operation (both physical and statistical), AD is vital for finding optimal parameter of quantum circuits. It is also required for feature map differentiation and solving differential equations~\cite{Kyriienko2021}. Similar strategies are also needed for Hamiltonian differentiation~\cite{OBrien2019, Mitarai2020}.

Analytic derivatives of quantum circuits can be estimated by measuring overlaps between quantum states, and were proven to benefit circuit optimization~\cite{Harrow2019}. This strategy is valid for differentiation of unitaries $\hat{U}(x) = \exp(-i x \hat{G}/2)$ generated by arbitrary Hermitian operators $\hat{G}$ --- a \emph{generator}. At the same time, an overlap measurement is different from expectations, as it requires a SWAP or Hadamard test that increase both circuit depth and system size~\cite{Ekert2002,Higgott2019}. To offer a resource-frugal quantum AD, the \emph{parameter shift rule} (PSR) was proposed and analyzed~\cite{Mitarai2018,Schuld2019,Li2017}. This algorithm is designed to estimate an analytic (exact, zero-bias) gradient via the measurement of expectation values, with quantum gate parameters being shifted to different values (originally considered to be fixed). However, the parameter shift rule is valid only for the specific type of generators~\cite{Mitarai2019a} that are involutory (i.e. $\hat{G}^2 = \mathbb{I}$ such that Euler's identity holds) or idempotent operators ($\hat{G}^2 = \hat{G}$, for example projectors). In this case the full analytic derivative requires only 2 measurements of expectations (corresponding to averaging over multiple shots for two distinct circuits and the same cost operator). The rules work for no more than two unique eigenvalues in the spectrum $\mathtt{diag}@\hat{G} \rightarrow \Lambda = \{ \lambda_j \}_{j=1}^{d}$, where $d = \mathrm{dim}(\hat{G})$. Namely, for involutory and idempotent generators we can shift the spectrum such that only $\{\pm \lambda \}$ eigenvalues appear (possibly repeated), leading to scaled identity $\hat{G}^2 = \lambda^2 \mathbb{I}
$~\cite{Schuld2019} that allows simplifying the differentiation. A similar approach was used for variational quantum chemistry protocols where generators in fermionic basis are split in parts with two multiply-degenerate eigenvalues that can be differentiated~\cite{Kottmann2021}.
PSR became \textit{de-facto} a go-to tool for QML algorithms, also implemented experimentally~\cite{Huang2020,Mari2021b}. This posed the quest for improving and generalizing the differentiation methods. The shifting rules were generalized to higher-order derivatives~\cite{Mitarai2020,Cerezo2021b,Mari2021} and arbitrary size yet symmetric shifts~\cite{Mari2021} that allow tuning a variance of the derivative estimation with zero bias (analytical at infinite number of shots). The generalization to non-symmetric parameter shifts and higher-order derivatives has just been proposed~\cite{Hubregtsen2021}. At the same time, generalizations beyond aforementioned generators remain limited. An important contribution was made in \cite{Vidal2018} where the Fourier transform approach was applied to differentiate multi-eigenvalue unitaries and perform coordinate descent. Also, an extension to the stochastic parameter shift rule with a composite two-term generator was proposed in \cite{Banchi2021}. For generic unitaries the circuit decomposition approach was proposed~\cite{Crooks2019}, being favourable in some cases, but not applicable to cases where hardware native generators have more than two eigenvalues.

In this paper we propose an approach that allows differentiating generic quantum circuits with unitaries generated by operators with a rich spectrum. The generalized differentiation rules are based on a spectral decomposition and show that a decisive role in differentiation is played by eigenvalue differences --- spectral gaps --- for the generator spectrum. Unlike the parameter shift rule valid for involutory and idempotent operators with a single unique spectral gap, our approach works for multiple non-degenerate gaps. We describe a workflow for derivative measurement based on: 1) analyzing the generator's spectrum; 2) calculating coefficients based on gaps and shifts; 3) measuring function expectations at shifted parameters (see Fig.~\ref{fig:workflow} for the visualization). We showcase the power of the approach using single- and two-qubit gates. Specifically, we show that the \texttt{fSim} gate (native to some superconducting circuits) can be differentiated with the proposed generalized rules without prior circuit decomposition. We also analyse variances for the derivative measurements estimated at finite number of shots. Our work provides a toolbox crucial for optimization and feature differentiation for digital-analog protocols and hardware-native QML.


\section{Method}

\subsection{Preliminaries}

We start by introducing a function encoding using the parametrized quantum circuit. For this we employ a variational state $ |\Psi_{\bm{\theta}}\rangle := \hat{V}_{\bm{\theta}}|{\O}\rangle$ that is prepared by a sequence of quantum gates (or Hamiltonian evolution) $\hat{V}_{\bm{\theta}}$ that depends on the vector of $K$ parameters, $\bm{\theta} \in \mathbb{R}^K$. Here, $|{\O}\rangle$ is a suitable initial state, typically taken as a product state. To read out the function value we also use the cost operator $\hat{\mathcal{C}}$ (this also can be a sum of Hermitian operators), such that the full expression reads $f_{\bm{\theta}} = \langle \Psi_{\bm{\theta}}| \hat{\mathcal{C}}| \Psi_{\bm{\theta}} \rangle$. Having the list of variational parameters $\{ \theta_k \}_{k=1}^{K}$, we are often interested in derivatives with respect to individual parameters. It is thus convenient to single out the parameter of interest, which we denote as $x$, while the vector of remaining parameters is labeled as $\widebar{\bm{\theta}}$. The choice of $x$ is arbitrary, and thus discussion remains general. We split the variational circuit into a product of circuits as $\hat{V}_{\bm{\theta}} := \hat{W}_{\widebar{\bm{\theta}}} \hat{U}(x) \hat{V}_{\widebar{\bm{\theta}}}$, and absorb a first part of the variational circuit into the initial state, $|\psi_{\widebar{\bm{\theta}}} \rangle := \hat{V}_{\widebar{\bm{\theta}}} |{\O}\rangle$, and introduce the dressed cost operator as $\hat{\mathcal{C}}_{\widebar{\bm{\theta}}} = \hat{W}^{\dagger}_{\widebar{\bm{\theta}}} \hat{\mathcal{C}} \hat{W}_{\widebar{\bm{\theta}}}$. The $x$-dependent unitary $\hat{U}(x)$ to be differentiated can be written in the general form as
\begin{align}
\label{eq:Ux}
\hat{U}(x) = \exp\left(-i \frac{\varphi(x)}{2} \hat{G}\right),
\end{align}
where $\hat{G}$ is a generator (Hermitian operator), and $\varphi(x)$ is a generic function, possibly nonlinear, of the encoded variable. In the following for simplicity we use $\varphi(x) = x$, and discuss the generalization towards the end of the section. 
\begin{figure}[t]
\begin{center}
\includegraphics[width=1.0\linewidth]{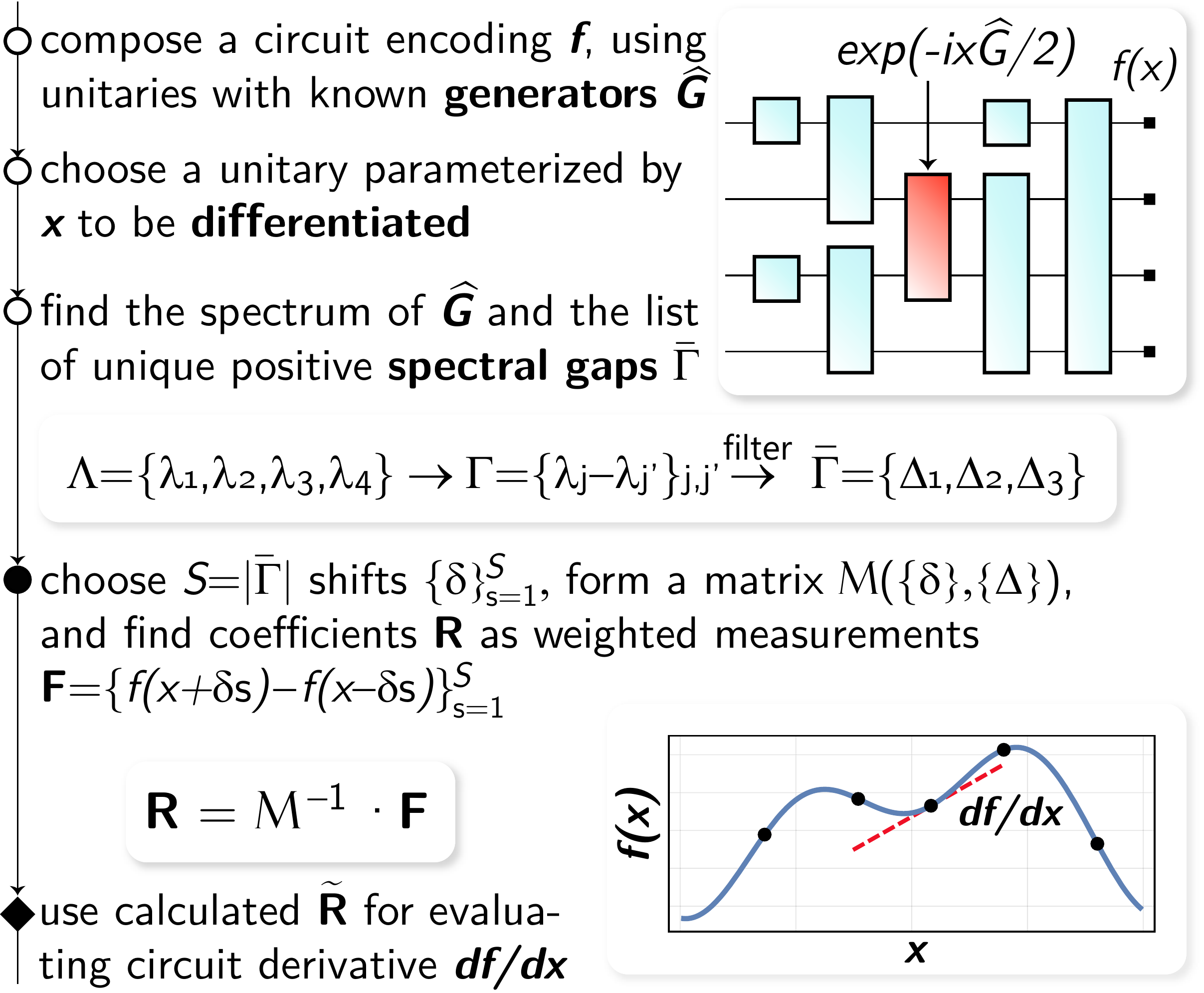}
\end{center}
\caption{\textbf{Workflow for generalized circuit differentiation.} Given a quantum circuit encoding a function $f$, we pick a unitary $\hat{U}(x)=\exp(-ix\hat{G}/2)$ parameterized by $x$, and study the spectrum of its generator $\hat{G}$. Using unique positive (non-zero) spectral gaps, we form a system of equations that includes spectral information, parameter shifts, and measured function expectations. The solution of this system allows estimating the analytical derivative for general $\hat{G}$.}
\label{fig:workflow}
\end{figure}
Using the definition presented above we are ready to present a function $f$ as an expectation value
\begin{align}
\label{eq:func}
f(x) = \langle \psi_{\widebar{\bm{\theta}}} | \hat{U}^{\dagger}(x) \hat{\mathcal{C}}_{\widebar{\bm{\theta}}} \hat{U}(x) | \psi_{\widebar{\bm{\theta}}} \rangle = \langle {\O} | \hat{V}_{\widebar{\bm{\theta}}}^{\dagger} e^{+i \frac{x}{2} \hat{G}} \hat{W}_{\widebar{\bm{\theta}}}^{\dagger} \hat{\mathcal{C}} \hat{W}_{\widebar{\bm{\theta}}} e^{-i \frac{x}{2} \hat{G}} \hat{V}_{\widebar{\bm{\theta}}} | {\O} \rangle,
\end{align}
where for brevity we do not state the $\widebar{\bm{\theta}}$-dependence for $f$. Note that the function representation in Eq.~\eqref{eq:func} can be seen through the prism of adjustable parametrized quantum circuits as in QNN or VQA architectures, as well as the differentiation of quantum feature maps~\cite{Kyriienko2021}. Straightforward differentiation of Eq.~\eqref{eq:func} with respect to $x$ yields
\begin{align}
\label{eq:dfdx}
\frac{d f(x)}{dx} = \frac{i}{2} \langle \psi_{\widebar{\bm{\theta}}} | e^{+i \frac{x}{2} \hat{G}} \left[ \hat{G}, \hat{\mathcal{C}}_{\widebar{\bm{\theta}}} \right] e^{-i \frac{x}{2} \hat{G}} | \psi_{\widebar{\bm{\theta}}} \rangle.
\end{align}
Eq.~\eqref{eq:dfdx} indeed holds for generic unitaries $\hat{U}(x)$ and generators $\hat{G}$, but has the form of an \emph{overlap}, and not an \emph{expectation value}. The parameter shift rule was proposed to rewrite Eq.~\eqref{eq:dfdx} as a central difference of two circuit evaluations valid for $\hat{G}^2 = \lambda^2 \mathbb{I}
$~\cite{Schuld2019}. This is valid for involutory and idempotent generators, where the spectrum with filtered duplicates is $\{ \pm \lambda \}$. In the following, we take a different route and generalize the rules to generators with more rich spectrum.


\subsection{Generalized circuit differentiation through spectral decomposition}

To develop the rules for differentiation we use the spectral decomposition for the generator. Under a suitable unitary transformation $\hat{\mathcal{U}}_{G}$ the generator can be diagonalized as
\begin{align}
\label{eq:UGU}
\hat{\mathcal{U}}_{G}^\dagger \hat{G} \hat{\mathcal{U}}_{G} = \sum\limits_{j} \lambda_j \hat{\mathcal{P}}_j =: \hat{D}, 
\end{align}
where $\hat{\mathcal{P}}_j \equiv |j\rangle \langle j|$ is a projector on the $j$th eigenstate of $\hat{G}$. In principle, we can always use diagonal generators, and use additional unitaries to change the basis, $\exp(-ix\hat{G}/2) = \hat{\mathcal{U}}_{G} \exp(-ix\hat{D}/2) \hat{\mathcal{U}}_{G}^\dagger$. Crucially, compared to the previously considered involutory and idempotent generators [for instance, sums of projectors satisfying $(\sum_{j} \hat{\mathcal{P}}_j)^2 = \sum_{j} \hat{\mathcal{P}}_j$], diagonal generators do not square to (scaled) identity, $\hat{D}^2 = \sum_{j} \lambda_j^2 \hat{\mathcal{P}}_j$. At the same time, they offer an easy access to matrix functions, where any function $g(\cdot)$ can be written in the form of weighted projectors, $g(\hat{D}) = \sum_j g(\lambda_j) \hat{\mathcal{P}}_j$. 
Using the spectral decomposition (or simply diagonal generators) we can rewrite Eq.~\eqref{eq:dfdx} as
\begin{align}
\label{eq:dfdx_diag}
\frac{df(x)}{dx} &= \frac{d}{dx} \langle \psi_{\widebar{\bm{\theta}}} | \Big( \sum\limits_{j=1}^{d} e^{+i \frac{x}{2} \lambda_{j}} \hat{\mathcal{P}}_{j} \Big) \hat{\mathcal{C}}_{\widebar{\bm{\theta}}} \Big( \sum\limits_{j'=1}^{d} e^{-i \frac{x}{2} \lambda_{j'}} \hat{\mathcal{P}}_{j'} \Big) | \psi_{\widebar{\bm{\theta}}} \rangle = \\ \notag
& = \frac{i}{2} \sum\limits_{j,j'=1}^{d} (\lambda_{j} - \lambda_{j'}) e^{i \frac{x}{2} (\lambda_{j} - \lambda_{j'})} \langle \psi_{\widebar{\bm{\theta}}} | \hat{\mathcal{P}}_{j}  \hat{\mathcal{C}}_{\widebar{\bm{\theta}}} \hat{\mathcal{P}}_{j'} | \psi_{\widebar{\bm{\theta}}} \rangle,
\end{align}
where we perform differentiation and show that the function derivative is composed of unknown matrix elements of the projected cost operator weighted by (complex) coefficients. The double summation in Eq.~\eqref{eq:dfdx_diag} introduces multiple terms. However, we find that by accounting for their properties we can reduce their number. 
We consider eigenvalues $\Lambda$ to be ordered, $\lambda_1 \geq \lambda_2 \geq ... \geq \lambda_d$. Next, we split Eq.~\eqref{eq:dfdx_diag} into three groups of terms. The first groups includes terms with $j<j'$. The second group includes $j>j'$ terms, where each term weighted by the eigenvalue differences $\lambda_{j}-\lambda_{j'} \equiv -(\lambda_{j'}-\lambda_{j})$ changes the sign, as compared to the first group. Similarly, complex exponents $\exp[ix(\lambda_{j}-\lambda_{j'})/2]$ are conjugated. Crucially, we observe that for Hermitian operators $\langle \hat{\mathcal{P}}_{j}  \hat{\mathcal{C}}_{\widebar{\bm{\theta}}} \hat{\mathcal{P}}_{j'} \rangle = \langle \hat{\mathcal{P}}_{j'}  \hat{\mathcal{C}}_{\widebar{\bm{\theta}}} \hat{\mathcal{P}}_{j} \rangle^*$, where averaging $\langle \dots \rangle$ is over the state $| \psi_{\widebar{\bm{\theta}}} \rangle$. Finally, in the third group we formally collect the diagonal terms with $j = j'$, and it vanishes trivially for the derivative.

The sum in RHS of Eq.~\eqref{eq:dfdx_diag} yields a real number, as expected from the derivative of a real function. Note that terms involving degenerate eigenvalues $j\neq j'$ also do not contribute to the derivative. 
Thus, from the full spectrum $\Lambda$ and the associated list of \emph{differences} of eigenvalues $\Gamma := \{ \lambda_{j}-\lambda_{j'} \}_{j,j' = 1}^{d}$ --- \emph{spectral gaps} --- we need to filter only unique positive non-zero gaps ($j > j'$). We denote the filtered set as $\widebar{\Gamma}$. It is convenient to introduce a superindex $s(j,j')$ (s.t. $j \neq j'$) that labels spectral gaps $\widebar{\Gamma} = \{ \Delta_{s} \}_{s = 1}^{S}$, where $S = |\widebar{\Gamma}|$ is the cardinality of the filtered set, and we denote \emph{unique gaps} as $\Delta_{s(j,j')} \equiv |\lambda_j - \lambda_{j'}|$ for $\lambda_j \neq \lambda_{j'}$. The matrix elements are labeled as $\langle \hat{\mathcal{P}}_{j}  \hat{\mathcal{C}}_{\widebar{\bm{\theta}}} \hat{\mathcal{P}}_{j'} \rangle =: \mathcal{O}_{s}$ and $\langle \hat{\mathcal{P}}_{j'}  \hat{\mathcal{C}}_{\widebar{\bm{\theta}}} \hat{\mathcal{P}}_{j} \rangle = \mathcal{O}_{s}^*$, and we note that for degenerate gaps matrix elements are summed up. Using the introduced notation, we can write the derivative simply as
\begin{align}
\label{eq:dfdx_simp}
\frac{df(x)}{dx} = \sum\limits_{s=1}^{S} \Delta_s R_s,
\end{align}
where we defined the real coefficients $\{ R_s \}_{s=1}^{S}$ such that
\begin{align}
\label{eq:Rs}
R_s &= \frac{i}{2} e^{i x \Delta_s/2} \mathcal{O}_{s} - \frac{i}{2} e^{-i x \Delta_s/2} \mathcal{O}_{s}^* \\ \notag 
&= -\sin\left( \frac{x}{2} \Delta_s \right) \mathrm{Re}\{\mathcal{O}_s\} -\cos\left( \frac{x}{2} \Delta_s \right) \mathrm{Im}\{\mathcal{O}_s\}.
\end{align}
We note there is at most $S_{\mathrm{max}} = d (d-1)/2$ unique coefficients, and if some spectral gaps in $\Gamma$ are degenerate this number is reduced.
Finding the associated vector $\mathbf{R}$ of coefficients is thus required for estimating the derivative \eqref{eq:dfdx_simp}. Next, we show how to access them through measurement of expectations $f(x + \delta)$ for shifted parameters.

Let us evaluate the function with a shifted argument 
\begin{align}
\label{eq:fx_delta}
f(x + \delta) =  \sum\limits_{s=1}^{S} \exp\left[i \frac{\Delta_{s}}{2} (x + \delta)\right] \mathcal{O}_{s} + \mathrm{h.c.},
\end{align}
where in the second equality we used the expression for higher-order derivative $f^{(n)} (x) \equiv d^{n} f/dx^{n}$,
\begin{align}
\label{eq:dfdx_n}
f^{(n)}(x) = \sum\limits_{s = 1}^{S} \left\{ \left(\frac{i\Delta_s}{2}\right)^{n} e^{i x \Delta_s/2} \mathcal{O}_{s} + \left(-\frac{i\Delta_s}{2}\right)^{n} e^{i x \Delta_s/2} \mathcal{O}_{s}^{*} \right\},
\end{align}
and collected the expansion in $n$, leading to exponent $\sum_n (i/2)^n \Delta_{s}^n \delta^n / n! = \exp[i \Delta_{s} \delta/2]$. Note that in Eq.~\eqref{eq:fx_delta} for brevity we omitted the constant term originating from $j=j'$ matrix elements, which vanishes for all derivatives except of $f^{(0)}(x)$ term. Using different shift strategies, we are now in a position to search for $\mathbf{R}$, and thereby writing explicit expressions for $df/dx$.\vspace{2mm}

\textit{Symmetric shifts.---}First, we consider the case of symmetric shifts. Namely, we aim to perform pairs of shifts, with plus and minus sign, keeping the magnitude of the shifts the same. This leads directly to evaluation through real coefficients $R_s$, avoiding the search for complex matrix elements $\mathcal{O}_s$. Let us consider a specific $\delta$ and two function evaluations. Subtracting minus-shifted from plus-shifted function measurement, we get
\begin{align}
\notag
f(x+\delta) - f(x-\delta) &= \sum\limits_{s=1}^{S} \left(e^{i\frac{\delta}{2}\Delta_s} - e^{-i\frac{\delta}{2}\Delta_s} \right) e^{i\frac{x}{2}\Delta_s} \mathcal{O}_s + \mathrm{h.c.} \\ &= 4 \sum\limits_{s=1}^{S} \sin\left(\frac{\delta \Delta_s}{2} \right) R_s,
\label{eq:f_diff}
\end{align}
where we used the same definitions as in Eq.~\eqref{eq:Rs}. Having established in Eq.~\eqref{eq:f_diff} the relation between real coefficients and central differences that are valid for any $\delta$, we need to choose a set of shifts $\{ \delta_s\}_{s=1}^{S}$ (corresponding to $2S$ measurements as we use $\pm \delta_s$ evaluations), and solve the system for $S$ values of $\{ R_s \}$. We can loosely call these shifted points as stencils that are used to improve the derivative quality~\cite{Bespalova2021}. Denoting central differences as $F_{s} := f(x+\delta_s) - f(x-\delta_s)$, the system of equations reads
\begin{align}
\label{eq:system}
    \begin{cases}
      F_1 = 4 \sum\limits_{s=1}^{S} \sin\left(\frac{\delta_1 \Delta_s}{2} \right) R_s,\\
      F_2 = 4 \sum\limits_{s=1}^{S} \sin\left(\frac{\delta_2 \Delta_s}{2} \right) R_s,\\
      \hspace{13.2mm}\cdots \\
      F_S = 4 \sum\limits_{s=1}^{S} \sin\left(\frac{\delta_S \Delta_s}{2} \right) R_s.\\
    \end{cases}\,
\end{align}
Introducing a vector of measured central differences $\mathbf{F} = \left( F_1, F_2, \dots, F_S \right)$, the system of Eqs.~\eqref{eq:system} recasts as $\mathbf{F} = \mathbb{M} \cdot \mathbf{R}$, where $\mathbb{M}$ is the matrix formed by sine functions evaluated for known phases. Inverting the system we can get the unknown coefficients as
\begin{align}
\label{eq:system_mat}
\mathbf{R} = \mathbb{M}^{-1} \cdot \mathbf{F}.
\end{align}
These coefficients can now be measured and evaluated as $R_s = \sum_{s'}(\mathbb{M}^{-1})_{s s'} F_{s'}$,
and substituting into Eq.~\eqref{eq:dfdx_simp} we get the \emph{measured} first-order derivative. While being simple, the approach holds for generic generators and represents one of the main results of this paper. Note that the measured derivative is bias-free in the same sense as the parameter shift rule for specific generators --- it is exact when measured with infinite number of shots. It does not require physically changing the basis to the diagonal one --- this step is only needed to derive the coefficients in front of measured function expectations.

We summarize the workflow for symmetric generalized circuit differentiation in Fig.~\ref{fig:workflow}. Its main steps are: 1) find unique non-zero gaps in the generator spectrum; 2) form the inverse matrix with complex coefficients; 3) combine to represent the measured derivative.\vspace{2mm}

\textit{Distinct shifts: Triangulation approach.---}In principle, we can also use distinct shifts that are general and are not equal in absolute values. Let us consider a system of equations formed by functions evaluated at different shifts. For any two stencil values $\delta_{\ell}$ and $\delta_{\ell'}$ we have
\begin{align}
\notag
&f(x+\delta_{\ell}) - f(x+\delta_{\ell'}) = \sum\limits_{s=1}^{S} (e^{i\frac{\delta_{\ell}}{2}\Delta_s} - e^{i\frac{\delta_{\ell'}}{2}\Delta_s})  e^{i\frac{x}{2}\Delta_s} \mathcal{O}_s + \mathrm{h.c.} = \\ \notag
&= 4 \sum\limits_{s=1}^{S} \sin\left[ \frac{(\delta_{\ell'} - \delta_{\ell})}{4} \Delta_s \right] \Bigg( \sin\left[ \frac{(2x+ \delta_{\ell'} + \delta_{\ell})}{4} \Delta_s \right] \mathrm{Re}\{\mathcal{O}_s\} + \\ &~~+ \cos\left[ \frac{(2x+ \delta_{\ell'} + \delta_{\ell})}{4} \Delta_s \right] \mathrm{Im}\{\mathcal{O}_s\} \Bigg),
\label{eq:f_diff_gen}
\end{align}
where we observe that the real part in RHS of Eq.~\eqref{eq:f_diff_gen} depends on both real and imaginary part of matrix elements $\mathrm{Re}\{\mathcal{O}_s\}$ and $\mathrm{Im}\{\mathcal{O}_s\}$. To determine these matrix elements we need to write the system of equations for $\ell, \ell' = 1, ..., 2 S + 1$ with independent function evaluations. 
This approach is directly related to Fourier-based circuit derivative introduced in \cite{Vidal2018} that requires the same number of measurements.
Importantly, to get derivatives for generic dressed cost function and initial state we need to have a reference point, which can be the function evaluated as zero shift. The naive approach based on $2S$ function evaluations fails to provide derivatives. The origin of this effect can be traced to the recently proven no-go theorem in \cite{Hubregtsen2021}.
Solving the system for $\mathrm{Re}\{\mathcal{O}_s\}$ and $\mathrm{Im}\{\mathcal{O}_s\}$, we substitute them in Eq.~\eqref{eq:Rs} and Eq.~\eqref{eq:dfdx_simp}, thus measuring the derivative with generalized shifts. Although the procedure is relatively straightforward, in most cases it resorts to numerical solution. While symmetric shifting resembles the central differencing in numerical differentiation (though exact), the generalized shifting procedure reminds of forward and backward differencing (also exact). We note that solving the generalized system may be beneficial in some cases as it allows playing with variances for the measured derivative.


\subsection{Feature map differentiation through spectral decomposition}

While circuit differentiation is used predominantly for optimizing variational parameters~\cite{Mitarai2018,Schuld2019,Hubregtsen2021}, recently it also became crucial for solving differential equations~\cite{Kyriienko2021}. In this case, $f(x)$ in Eq.~\eqref{eq:func} encodes a function of variable $x$, and $\hat{U}(x)$ serves as a \emph{feature map}~\cite{SchuldPRL}. Our goal is representing $df(x)/dx$ as a sum of expectation values, using the smallest number of evaluations.

We now show that for some feature maps and circuits used in QML we can use symmetries to reduce the number of circuit evaluations. Typically, feature maps represent a tensor product of rotations (parallel sequences) parameterized by the same variable $x$~\cite{Goto2020}. One example corresponds to the \textit{product feature map} that we define as~\cite{Kyriienko2021}
\begin{align}
\label{eq:feature_QCL}
\hat{U}(x) = \bigotimes_{j=1}^{N} \hat{R}_{\alpha,j} [\varphi(x)],
\end{align}
where $N$ is the number of qubits used for the encoding. $\hat{R}_{\alpha,j}(\varphi) := \exp\left(-i \frac{\varphi}{2} \hat{P}_{\alpha,j} \right)$ is a Pauli rotation operator for Pauli matrices $\hat{P}_{\alpha,j} = \hat{X}_j$, $\hat{Y}_j$, or $\hat{Z}_j$ ($\alpha = x,y,z$, respectively) acting on qubit $j$ with phase $\varphi$. For brevity, let us assume $\alpha = z$, as other cases can be considered in the same way with additional Hadamard and phase gates before and after the feature map. 

We can rewrite Eq.~\eqref{eq:feature_QCL} in the form of Eq.~\eqref{eq:Ux}, where the unitary operator is generated by
\begin{align}
\label{eq:G_product}
\hat{G}_{\mathrm{Z}} = \sum_{j=1}^{N} \hat{Z}_j
\end{align}
that corresponds to a sum of Pauli operators. Being the total effective magnetization~\cite{Kozin2019}, the generator $\hat{G}_{\mathrm{Z}}$ in Eq.~\eqref{eq:G_product} has a spectrum with $N+1$ eigenvalues $\Lambda = \{ -N, -N + 2, ..., N-2, N \}$. Some eigenvalues are multiply degenerate, and the number of unique positive gaps is $S= |\widebar{\Gamma}| = N$. As the feature map can also be differentiated by applying the parameter shift rule individually to each rotation, the derivative can be measured using at most $2N$ circuit evaluations.

First, we note that eigenstates of $\hat{G}_{\mathrm{Z}}$ coincide with computational basis states and have all real amplitudes. Equivalently, these can be states where a global phase is the same. Next, we consider an input state $|{\O}\rangle$ with real amplitudes, and ansatze $\hat{V}_{\bm{\theta}}$, $\hat{W}_{\bm{\theta}} \in \mathcal{SO}(2^N)$ that preserve this structure (one example is used in fermionic simulations~\cite{Gard2019}), such that $|\Psi_{\bm{\theta}}\rangle$ remains real. Finally, we consider cost functions with real-valued matrix elements, which can be formed by strings of Pauli X and Z operators, as well as strings with the even number of Pauli Y operators. Given the real structure of (variational) states and readout, the matrix elements $\mathcal{O}_s \in \mathbb{R}$ in Eq.~\eqref{eq:f_diff_gen} are real for any set of shifts $\{ \delta_\ell \}$, and the difference considered in Eq.~\eqref{eq:f_diff_gen} now reads
\begin{align}
\notag
&f(x+\delta_{\ell}) - f(x+\delta_{\ell'}) = 4 \sum\limits_{s=1}^{S} \sin\left[ \frac{(\delta_{\ell'} - \delta_{\ell})}{4} \Delta_s \right] \times \\ & \times \sin\left[ \frac{(2x+ \delta_{\ell'} + \delta_{\ell})}{4} \Delta_s \right] \mathrm{Re}\{\mathcal{O}_s\},
\label{eq:f_diff_real}
\end{align}
and we remind that the imaginary parts of matrix elements $\mathrm{Im}\{ \mathcal{O}_s\}$ remain zero. As the number of unknown coefficients reduces by symmetry, it is sufficient to perform measurements at $N+1$ shifts (may include original function evaluation), reducing the budget of analytic derivative measurement from $2N$ evaluations required before for the single-qubit parameter shift rule. This halves the measurement time for function derivatives, and speeds up the solution for systems of differential equations.

So far we observed that the specific choices we made here can give us reductions in the number of required measurements. Similarly, we can also imagine the situation where only imaginary parts are non-zero, or the matrix elements are complex-valued but have specific structure that connects real and imaginary parts (like in Kramers–Kronig relations that may be familiar to physicists).  
To conclude the section, we note that many more symmetries can be incorporated to reduce the number of measurement. They depend on the structure of ansatze, feature maps, and cost functions, or can be facilitated by pre- and post-processing. Finding efficient strategies that minimize the number of function evaluations is an important direction of future research.


\section{Applications and Discussion}

In the previous section we established the rules for generic quantum circuit differentiation. Let us show how they can be applied in practice. As discussed above, exact formulae are defined by the generator gaps, and thus we consider possible generators of interest.

\subsection{Pauli string generators}

First, let us reconsider the case of involutory generators. These can be represented by Pauli operators (generating single-qubit rotations) or multi-qubit Pauli strings. Their spectrum is $\widebar{\Lambda} = \{ +1, -1\}$ (repeated eigenvalues are filtered out), the gap spectrum yields $\Gamma = \{0, 2, -2, 0\}$, and $\widebar{\Gamma} = \{ \Delta \}$, where $\Delta = 2$. Only single shift value suffices, where Eq.~\eqref{eq:f_diff} substituted to Eq.~\eqref{eq:dfdx_simp} leads to
\begin{align}
\label{eq:dfdx_1q}
\frac{df(x)}{dx} = \frac{\Delta [f(x+\delta) - f(x-\delta)] }{4 \sin (\delta \Delta /2)}.
\end{align}
This result thus coincides with other works with PSR generalizations in \cite{Mari2021,Hubregtsen2021}.

To use the generalized approach with distinct shift triangulation for spectra with the single unique gap we formally require three shifts $\delta_{1,2,3}$. Note that one of them may be zero, thus leading to a simple function evaluation. The generalized derivative for $S=1$ reads
\begin{align}
\label{eq:dfdx_1q_gen}
&\frac{df(x)}{dx} = \frac{\Delta}{8} \Bigg\{ \left[\cos\left(\frac{\delta_2 \Delta}{2} \right) - \cos\left(\frac{\delta_3 \Delta}{2} \right)\right] f(x + \delta_1) + \\ \notag &+ \left[\cos\left(\frac{\delta_3 \Delta}{2} \right) - \cos\left(\frac{\delta_1 \Delta}{2} \right)\right] f(x + \delta_2) + \Bigg[\cos\left(\frac{\delta_1 \Delta}{2} \right) - \\ \notag &- \cos\left(\frac{\delta_2 \Delta}{2} \right)\Bigg] f(x + \delta_3) \Bigg\} \Bigg\{ \sin\left[\frac{(\delta_2 - \delta_1) \Delta}{4} \right] \sin\left[\frac{(\delta_2 - \delta_3) \Delta}{4} \right] \\ \notag &\sin\left[\frac{(\delta_3 - \delta_1) \Delta}{4} \right] \Bigg\}^{-1}.
\end{align}
While the three-shift generalization shown in Eq.~\eqref{eq:dfdx_1q_gen} was not considered before, intriguingly, it correlates with the no-go theorem recently presented in Ref.~\cite{Hubregtsen2021}, showing that only two shifts of distinct magnitude cannot provide the derivative of the PSR type. The presented generalization offers high tunability in terms of variance minimization and maximal derivative inference, and shall be considered in cases where function evaluation $f(x)$ is anyway available (e.g. as in QML variational algorithms with $\mathrm{L}_2$ loss).

While the major goal of finding analytic derivatives is in reducing the bias to zero, another important issue is in minimizing a variance for the derivative measurement $df/dx \equiv f'$. This is defined as 
\begin{align}
\label{eq:var_def}
\mathrm{Var}[f'] = \mathbb{E}\Big[(f')^2\Big] - \mathbb{E}\Big[f'\Big]^2,
\end{align}
where we remind that the measurement of the derivative assumes taking expectation value over $N_{\mathrm{shots}}$ shots. Following the approach by Mari \textit{et al.}~\cite{Mari2021}, we can derive the variance for derivative measurement assuming each function measurement obeys normal statistics with variance $\sigma^2(x)/N_{\mathrm{shots}}$. For the symmetric differentiation rule in Eq.~\eqref{eq:dfdx_1q} the derivative variance reads
\begin{align}
\label{eq:var_1q}
\mathrm{Var}[f'] = \frac{\Delta^2 }{16 \sin^2(\delta \Delta / 2)} \frac{\Big[\sigma^2(x + \delta) + \sigma^2(x - \delta) \Big]}{N_{\mathrm{shots}}},
\end{align}
and for the case of Pauli generators with $\Delta = 2$ this coincides with the result in Ref.~\cite{Mari2021}.
\begin{figure}[t]
\begin{center}
\includegraphics[width=1.0\linewidth]{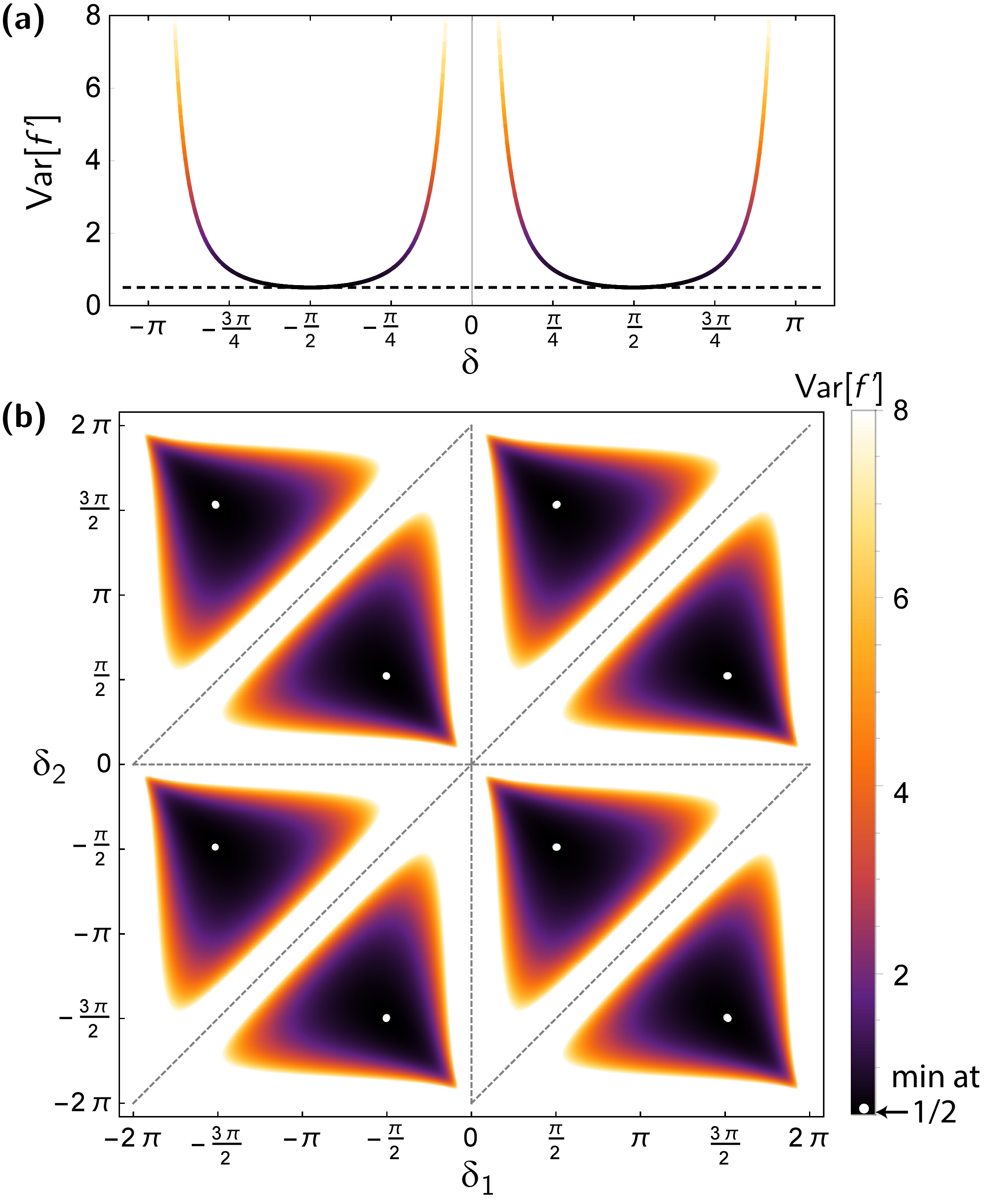}
\end{center}
\caption{\textbf{Variance for generalized circuit derivatives: $S=1$ case.} We consider generators with the single unique gap $\Delta = 2$. \textbf{(a)} Variance $\mathrm{Var}[f']$ shown as a function of the symmetric $\delta$ shift. The plot assumes that the variance of measured observables is shift-independent, and $\mathrm{Var}[f']$ is in units of $\sigma_0^2/N_{\mathrm{shots}}$. The minimal values of $1/2$ is shown by the dashed horizontal line. \textbf{(b)} Variance for derivative measured with distinct shifts $\delta_1$ and $\delta_2$. This is minimized at $1/2$ for combinations of $\pm \pi/2$ and $\pm 3\pi/2$ shown by white dots. Measurement statistics is assumed to be shift-independent. Dashed gray lines depict shift parameters at which the variance increases as it approaches the finite difference limit, or when no meaningful information about derivative is gained.}
\label{fig:var_S1}
\end{figure}

In general, the variance of the function measurement depends on the shift value, as well as initial state and the dressed cost function, making the full analysis circuit-dependent. However, as we see in Eq.~\eqref{eq:var_1q}, the variance contributions from $+\delta$ and $-\delta$ circuit cannot be tuned independently, and $\delta$ mainly controls the overall prefactor of $\mathrm{Var}[f']$. Using a simplistic assumption that the measurement statistics has only a weak parameter dependence, $\sigma^2(x) \approx \sigma_0^2$, one can write the expression for the variance of derivative measurement as $\mathrm{Var}[f'] \approx \Delta^2 \sigma_0^2 / [8 \sin^2(\delta \Delta/2) N_{\mathrm{shots}}]$. For illustration, we provide this dependence in Fig.~\ref{fig:var_S1}(a). The variance is measured in $\sigma_0^2 / N_{\mathrm{shots}}$ units, and is minimized at $1/2$ for $\pm \pi /2$ shifts.

The situation changes qualitatively when we consider the generalized $S=1$ circuit differentiation with the triangulation approach. Unlike the symmetric case, the derivative measurement in Eq.~\eqref{eq:dfdx_1q_gen} includes the linear combination from three circuit measurements. Each measurement is normally distributed due to sampling statistic, but may have different variance, and has a distinct prefactor. The variance for generalized derivative measurement is
\begin{align}
\label{eq:var_1q_gen}
\mathrm{Var}[f'] = \frac{\nu_1 \sigma^2(x + \delta_1) + \nu_2 \sigma^2(x + \delta_2) + \nu_3 \sigma^2(x + \delta_3)}{N_{\mathrm{shots}}},
\end{align}
where weighting coefficients $\nu_{1,2,3}$ are defined as
\begin{align}
\label{eq:nu1}
\nu_1 = \frac{\Delta^2 \Big[ \cos(\delta_2 \Delta/2) - \cos(\delta_3 \Delta/2) \Big]^2 }{64 \sin^2\Big[ \frac{(\delta_2 - \delta_1) \Delta}{4} \Big] \sin^2\Big[ \frac{(\delta_2 - \delta_3)\Delta}{4} \Big] \sin^2\Big[ \frac{(\delta_3 - \delta_1) \Delta}{4} \Big]}, 
\end{align}
\begin{align}
\label{eq:nu2}
\nu_2 = \frac{\Delta^2 \Big[ \cos(\delta_3 \Delta/2) - \cos(\delta_1 \Delta/2) \Big]^2 }{64 \sin^2\Big[ \frac{(\delta_2 - \delta_1) \Delta}{4} \Big] \sin^2\Big[ \frac{(\delta_2 - \delta_3)\Delta}{4} \Big] \sin^2\Big[ \frac{(\delta_3 - \delta_1) \Delta}{4} \Big]},
\end{align}
\begin{align}
\label{eq:nu3}
\nu_3 = \frac{\Delta^2 \Big[ \cos(\delta_1 \Delta/2) - \cos(\delta_2 \Delta/2) \Big]^2 }{64 \sin^2\Big[ \frac{(\delta_2 - \delta_1) \Delta}{4} \Big] \sin^2\Big[ \frac{(\delta_2 - \delta_3)\Delta}{4} \Big] \sin^2\Big[ \frac{(\delta_3 - \delta_1) \Delta}{4} \Big]}.
\end{align}
Importantly, the coefficients in Eqs.~\eqref{eq:nu1}--\eqref{eq:nu3} depend on three shifts, and potentially can be adjusted to minimize the full variance. Finally, to illustrate how the derivative variance depends on the shifting parameters, we set $\delta_3 = 0$ (simple function evaluation), assume $\sigma^2(x+\delta_\ell) \approx \sigma_0^2$ ($\ell=1,2,3$), and plot $\mathrm{Var}[f']$ as a function of $\delta_1$ and $\delta_2$. The result is shown in Fig.~\ref{fig:var_S1}(b). We observe several regions where the variance (measured in $\sigma_0^2/N_{\mathrm{shots}}$ units) is minimized at $1/2$ value. The corresponding shifts of $\pi/2$ and $3\pi/2$ magnitude are shown by white dots. We also note the regions of high variance $\mathrm{Var}[f']$ (white, cut at $\mathrm{Var}[f'] = 8$). These can be separated in two groups. First, for small shifts the variance increases similarly to numerical differentiation with finite differencing. Second, there are sets of shifts, depicted with dashed gray lines, where two shifts are equal, and thus no useful information about the derivative is used. To conclude this discussion, we stress that the optimization of shifts for generalized derivative measurement shall depend on $\sigma^2(x + \delta_\ell)$, $\ell=1,2,3$, and may contribute to the variance reduction using additional control tools.


\subsection{\texttt{fSim} gate}

Now let us consider an example of the gate where the capabilities of generalized parameter differentiation are crucial. This is represented by hardware-native unitaries that are generated by Hamiltonians with more than one unique spectral gap. While elementary gate decomposition can help to differentiate the sequence using derivative product rule~\cite{Crooks2019}, we note that for many hardware platforms it is beneficial to differentiate them without extra overheads (depth increase), and using already optimized structure. One example is the \texttt{fSim} gate native to Google's devices based on gmon architecture~\cite{Foxen2020}, which enabled state-of-the-art simulation of scrambling on a quantum processor~\cite{Mi2021}. This two-qubit gate is represented by the unitary parametrized by two angles $\theta$ and $\phi$ that reads
\begin{align}
\label{eq:UfSim_matrix}
  \hat{U}_{\mathrm{fSim}}(\theta, \phi) = \begin{pmatrix}
    1 & 0 & 0 & 0 \\
    0 & \cos(\theta) & -i\sin(\theta) & 0 \\
    0 & -i\sin(\theta) & \cos(\theta) & 0 \\
    0 & 0 & 0 & \exp(-i\phi)
  \end{pmatrix}.
\end{align}
This unitary can be formally written as $\hat{U}_{\mathrm{fSim}}(\theta, \phi) = \exp[-i \hat{\mathcal{H}}_{\mathrm{fSim}}(\theta, \phi)]$ where the underlying effective Hamiltonian can be deduced as a matrix logarithm $\hat{\mathcal{H}}_{\mathrm{fSim}}(\theta, \phi) = i \log [\hat{U}_{\mathrm{fSim}}(\theta, \phi)]$. This can be written in Pauli basis as
\begin{align}
\label{eq:HfSim_Pauli}
\hat{\mathcal{H}}_{\mathrm{fSim}}(\theta, \phi) = \frac{\theta}{2} \Big( \hat{X}_1 \hat{X}_2 + \hat{Y}_1 \hat{Y}_2 \Big) + \frac{\phi}{4} \Big( \mathbb{I} - \hat{Z}_1 - \hat{Z}_2 + \hat{Z}_1 \hat{Z}_2 \Big).
\end{align}
Following the notation introduced in the previous section, we assign a quantum circuit to be differentiated as
\begin{align}
\label{eq:UfSim}
\hat{U}_{\mathrm{fSim}} = \exp\left[ -i \frac{\theta}{2} \hat{G}^{(\theta)} -i \frac{\phi}{2} \hat{G}^{(\phi)} \right],
\end{align}
where we defined separate generators $\hat{G}^{(\theta)} := \left( \hat{X}_1 \hat{X}_2 + \hat{Y}_1 \hat{Y}_2 \right)$ and $\hat{G}^{(\phi)} := \left( \mathbb{I} - \hat{Z}_1 - \hat{Z}_2 + \hat{Z}_1 \hat{Z}_2 \right) / 2$. When the \texttt{fSim} gate is a part of the bigger circuit and function (or energy) encoding, its derivative with respect to $\theta$ reads
\begin{align}
\label{eq:dfdx_fSim}
\frac{d f(\theta, \phi)}{d\theta} = \frac{i}{2} \langle \psi_{\widebar{\bm{\theta}}} | e^{+i \left(\frac{\theta}{2} \hat{G}^{(\theta)} + \frac{\phi}{2} \hat{G}_{2}^{(\phi)} \right) } \left[ \hat{G}^{(\theta)}, \hat{\mathcal{C}}_{\widebar{\bm{\theta}}} \right] e^{-i \left(\frac{\theta}{2} \hat{G}^{(\theta)} + \frac{\phi}{2} \hat{G}^{(\phi)} \right) } | \psi_{\widebar{\bm{\theta}}} \rangle,
\end{align}
and the derivative with respect to $\phi$ can be written in a similar way. Following the spectral decomposition approach we need to bring the generator into diagonal form, $\hat{\mathcal{U}}_{G}^{\dagger} (\theta \hat{G}^{(\theta)} + \phi \hat{G}^{(\phi)} ) \hat{\mathcal{U}}_{G} =: \theta \hat{D}^{(\theta)} + \phi \hat{D}^{(\phi)}$. Here the diagonal $\theta$-generator reads $\hat{D}^{(\theta)} := \hat{Z}_2 - \hat{Z}_1$, while $\phi$-generator remains the same, $\hat{D}^{(\phi)} := \hat{G}^{(\phi)}$. Since $\hat{G}^{(\theta)}$ and $\hat{G}^{(\phi)}$ commute the basis-change unitary $\hat{\mathcal{U}}_{G}$ is parameter-independent, so the consideration can follow the same spectral decomposition strategy as in Eq.~\eqref{eq:dfdx_diag}. The derivative with respect to $\theta$ and $\phi$ can be analyzed studying generators $\hat{D}^{(\theta)}$ and $\hat{D}^{(\phi)}$. We start with $\theta$ differentiation as more interesting case.
\begin{figure}[t]
\begin{center}
\includegraphics[width=0.96\linewidth]{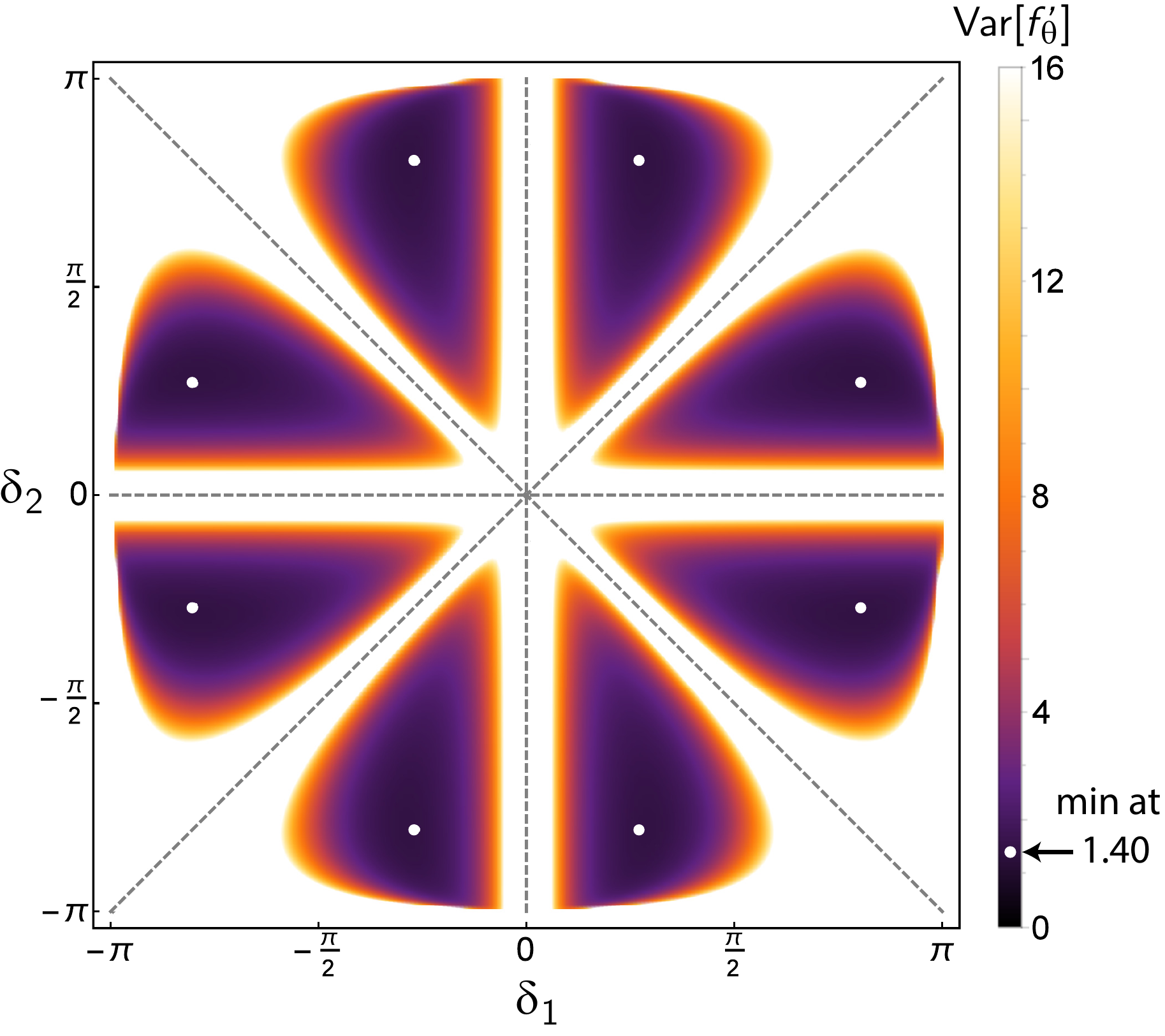}
\end{center}
\caption{\textbf{Variance for \texttt{fSim} gate derivative: $S=2$ case.} We consider \texttt{fSim} generator with two unique gap $\Delta_1 = 2$ and $\Delta_2 = 4$ corresponding to $\theta$ differentiation. Variance $\mathrm{Var}[f_{\theta}']$ is shown as a function of two symmetric shifts $\delta_1$ and $\delta_2$. It is measured in units of $\sigma_0^2/N_{\mathrm{shots}}$ assuming that function measurements are shift-independent. The optimal shifts are shown by white dots and are located around $\pm 0.8 \pi$ and $\pm 0.29 \pi$ values, leading to the minimal variance of about $1.4$. Dashed gray lines show cases where no useful information about the derivative is measured.
}
\label{fig:var_fSim}
\end{figure}

The spectrum of $\hat{D}^{(\theta)}$ (or equivalently $\hat{G}^{(\theta)}$) consists of four eigenvalues, $\Lambda^{(\theta)} = \{ 2, 0, 0, -2\} $ (same as for single-qubit rotations on two separate qubits). The filtered set of non-zero positive gaps includes $\widebar{\Gamma} = \{ \Delta_1, \Delta_2\} $, where $\Delta_1 = 2$ and $\Delta_2 = 4$. Thus, $\theta$ differentiation requires the $S=2$ strategy for finding two unknown coefficients. Using the symmetric shifting procedure with four measurements at $\{\pm \delta_1, \pm \delta_2\}$, we arrive at the explicit expression for the \texttt{fSim} derivative
%
%
%
\begin{align}
\label{eq:dfdtheta_fSim}
\frac{df(\theta, \phi)}{d\theta} &= \alpha_1(\delta_{1,2}, \Delta_{1,2})\Big[f(\theta +\delta_1, \phi) - f(\theta-\delta_1, \phi)\Big] + \\ \notag &~+ \alpha_2(\delta_{1,2}, \Delta_{1,2}) \Big[f(\theta +\delta_2, \phi) - f(\theta -\delta_2, \phi)\Big],
\end{align}
where we defined coefficients
\begin{align}
\label{eq:alpha_1}
\alpha_1 = \frac{\Delta_1 \sin \bigg( \frac{\delta_2 \Delta_2}{2} \bigg) -  \Delta_2 \sin \bigg( \frac{\delta_2 \Delta_1}{2}\bigg) }{4 \Bigg\{ \sin \bigg( \frac{\delta_1 \Delta_1}{2} \bigg) \sin \bigg( \frac{\delta_2 \Delta_2}{2} \bigg) - \sin \bigg( \frac{\delta_1 \Delta_2}{2} \bigg) \sin \bigg( \frac{\delta_2 \Delta_1}{2} \bigg)  \Bigg\}},
\end{align}
\begin{align}
\label{eq:alpha_2}
\alpha_2 = \frac{\Delta_2 \sin \bigg( \frac{\delta_1 \Delta_1}{2} \bigg) -  \Delta_1 \sin \bigg( \frac{\delta_1 \Delta_2}{2}\bigg) }{4 \Bigg\{ \sin \bigg( \frac{\delta_1 \Delta_1}{2} \bigg) \sin \bigg( \frac{\delta_2 \Delta_2}{2} \bigg) - \sin \bigg( \frac{\delta_1 \Delta_2}{2} \bigg) \sin \bigg( \frac{\delta_2 \Delta_1}{2} \bigg)  \Bigg\}},
\end{align}
and Eqs.~\eqref{eq:dfdtheta_fSim}-\eqref{eq:alpha_2} hold for any generator with two unique gaps. 
We note that the four-term parameter-shift rule was derived in \cite{Anselmetti2021} in the context of VQE, where a commutator-based approach was used. 
The triangulation approach can also be used using five measurements, with four distinct non-zero shifts. However, the expressions are cumbersome and we avoid presenting them here.

We return back to the $\phi$ differentiation case. Looking at the spectrum of $\hat{G}^{(\phi)}$ corresponding to a projector, $\Lambda^{(\phi)} = \{2, 0, 0, 0 \} $, we get $\widebar{\Gamma} = \{\Delta\}$, where $\Delta = 2$. Thus, only two function evaluations are needed following the $S=1$ rule in Eq.~\eqref{eq:dfdx_1q},
\begin{align}
\label{eq:dfdphi_fSim}
\frac{df(\theta, \phi)}{d\phi} = \frac{\Delta [f(\theta, \phi+\delta) - f(\theta, \phi -\delta)] }{4 \sin (\delta \Delta /2)},
\end{align}
or the generalized $S=1$ rule in Eq.~\eqref{eq:dfdx_1q_gen}.

Now, let say a couple of words about the variance. Since $\phi$ differentiation follows the $S=1$ rules (we consider symmetric shifts), the variance scaling is the same as for Pauli string generators considered before. However, the situation changes for the $\theta$ differentiation. The corresponding variance can be written as
\begin{align}
\label{eq:var_fSim}
\mathrm{Var}[f_{\theta}'] &= \alpha_1^2 \Big[\sigma^2(\theta + \delta_1,\phi) + \sigma^2(\theta - \delta_1,\phi) \Big] / N_{\mathrm{shots}} \\ \notag &+ \alpha_2^2 \Big[\sigma^2(\theta + \delta_2,\phi) + \sigma^2(\theta - \delta_2,\phi) \Big] / N_{\mathrm{shots}},
\end{align}
showing that $\delta_1$ and $\delta_2$ shifts come with distinct weights. Assuming for simplicity that the statistics for function measurements is the same, $\sigma^2(\theta \pm \delta_s,\phi) \approx \sigma_0^2$, we plot the derivative variance $\mathrm{Var}[f_{\theta}']$ in Fig.~\ref{fig:var_fSim}. We observe that the variance is minimized at shifts around $0.80 \pi$ and $0.29 \pi$ in absolute values (white dots), with $\mathrm{Var}[f_{\theta}']$ reaching $1.40$ in $\sigma_0^2/N_{\mathrm{shots}}$ units. Naturally, for equal shifts or for zero shifts we cannot access the derivative faithfully, as shown by increasing variance (dashed gray lines in Fig.~\ref{fig:var_fSim}).


\subsection{Other two-qubit gates}

Other examples of two-qubit gates that are based on multigap spectrum include the cross-resonance (\texttt{CR}) gate as realized in superconducting circuits~\cite{Chow2011}. The generator of \texttt{CR} gate corresponds to a sum of different Pauli strings. This sum generally corresponds to the effective Hamiltonian of the Hilbert space for microwave modes (non-linear bosonic ladders) truncated to lowest two levels. Due to the multilevel structure many terms arise with the strength being dependent on (and/or tuned by) the drive power, anharmonicity, cross-talk etc \cite{Sheldon2016,Magesan2020,Malekakhlagh2020}. The effective circuit Hamiltonian also can be modified by using Floquet engineering~\cite{Kyriienko2018}. The structure of the effective \texttt{CR} Hamiltonian in general corresponds to~\cite{Malekakhlagh2020}
\begin{align}
\label{eq:H_CR_gen}
\hat{\mathcal{H}}_{\mathrm{CR}} = \gamma_1 \hat{Z}_1 + \gamma_2 \hat{Z}_1 \hat{X}_2 + \gamma_3 \hat{X}_2 + \gamma_4 \hat{Z}_2 + \gamma_5 \hat{Z}_1 \hat{Z}_2,
\end{align}
where $\{\gamma_j\}$ are tunable parameters that dependent on the system, and additionally spurious $\sim \hat{Y}_2$ terms can appear due to cross-talk~\cite{Magesan2020}. Typically one can select dominant terms by tuning the drive power. Ideally, \texttt{CR} gate shall be generated by the $\hat{Z}_1 \hat{X}_2$ term. In this case, with the pure Pauli string generator, the $S=1$ parameter shift rule is sufficient for the gate differentiation. 

In some cases the cross-resonance drive generates strong single-qubit terms simultaneously with the two-qubit cross-resonant interaction. First, for the dominant $\hat{Z}_1$ and $\hat{Z}_1 \hat{X}_2$ terms of the same magnitude~\cite{Magesan2020}, $\gamma_1 = -\gamma_2 \equiv 1$, and $\gamma_{3,4,5}=0$, the \texttt{CR} gate generator reduces to $\hat{G}_{\mathrm{CR}} = \hat{Z}_1 - \hat{Z}_1 \hat{X}_2$ the spectrum is $\Lambda = \{2, 0, 0, -2 \}$ and gaps are $\widebar{\Gamma} = \{2, 4\}$. In this case one can use the four-shift rule described in Eq.~\eqref{eq:dfdtheta_fSim}.

In cases when the cross resonant drive induces single qubit terms of equal strength, $\gamma_1 = \gamma_3 = 1$, and the two-qubit \texttt{CR} interaction is weaker but non-zero, $-1< \gamma_2 < 0$ ~\cite{Magesan2020}, the generator corresponds to $\hat{G}_{\mathrm{CR}} = \hat{Z}_1 + \gamma_2 \hat{Z}_1 \hat{X}_2 + \hat{X}_2$, and leads to $\Lambda = \{ -2 + \gamma_2, \gamma_2, -\gamma_2, 2 + \gamma_2 \}$ spectrum. The corresponding filtered gaps include for three distinct values $\widebar{\Gamma} = \{\Delta_1, \Delta_2, \Delta_3 \}$ that include  $\Delta_1 = 2 (1 + \gamma_2)$, $\Delta_2 = 2 (1 - \gamma_2)$, and $\Delta_3 = 4$. For brevity, setting $\gamma_2 = -1/2$ we arrive at $\widebar{\Gamma} = \{ 1, 3, 4 \}$ distinct spectral gaps. To differentiate this gate we require six circuit evaluations in the symmetric case. The $S=3$ differentiation rule can be written as
\begin{align}
\label{eq:dfdx_CR}
\frac{df(x)}{dx} &= \frac{\nu_1}{4 \mathcal{V}}\Big[f(x +\delta_1) - f(x-\delta_1)\Big] + \\ \notag &+ \frac{\nu_2}{4 \mathcal{V}} \Big[f(x +\delta_2) - f(x-\delta_2)\Big] \\ \notag &+ \frac{\nu_3}{4 \mathcal{V}} \Big[f(x +\delta_3) - f(x-\delta_3)\Big],
\end{align}
where coefficients $\nu_{1,2,3}$ and the denominator $\mathcal{V}$ have involved trigonometric dependence on gaps $\Delta_{1,2,3}$ and shifts $\delta_{1,2,3}$. For completeness, we present the explicit expressions in Appendix \ref{A}.

For the situations with more Pauli terms included in the generator, and generic interaction constants $\gamma_j$, the spectrum is even more rich. It may include four distinct positive gaps. In this case differentiation with symmetric shifts requires $8$ function evaluations. The corresponding expressions are too bulky to be presented explicitly, and possibly the best course of action is numerical calculation of coefficients from Eq.~\eqref{eq:system_mat}.

Finally, for the generic two-qubit gates one can consider a generator as a sum of all possible Pauli strings. This gives us the maximal number of unique gaps $S=d(d-1)/2 = 6$ for the four-state Hilbert space. We have seen in the analysis of differentiation that this number matches the number of independent matrix elements $\{ \mathcal{O}_s \}$, with both real and complex parts, which for generic dressed cost function $\hat{\mathcal{C}}_{\widebar{\bm{\theta}}}$ and generic states $|\psi_{\widebar{\bm{\theta}}}\rangle$ are independent. We thus expect that the full information about the circuit derivative requires not less than $2S$ measurements, unless some of them vanish due to $\hat{\mathcal{C}}_{\widebar{\bm{\theta}}}$ or $|\psi_{\widebar{\bm{\theta}}}\rangle$ symmetries (as happens sometimes in practice).


\subsection{Qutrit gates}

As the generalization only depends on the spectrum, and is not tied to specific form of operators, we can also apply it to multilevel quantum systems. The minimal example here is a qutrit. The qutrit operators correspond to generalizations of Pauli matrices for three-level quantum systems, often taken as Gell-Mann matrices. In quantum hardware these are often realized in NV centers in diamond and atomic systems. From the application perspective, the ability to differentiate three-level unitary operators may largely benefit optimal control~\cite{Choi2020,Li2017}, and improve dynamical decoupling and sensing capabilities.

Possible generators of unitaries for qutrits as single SU(3) matrices include
\begin{align}
\label{eq:G_qutrit}
\hat{G}_{1}^{(3)} = \begin{pmatrix}
    0 & 1 & 0 \\
    1 & 0 & 0 \\
    0 & 0 & 0
  \end{pmatrix},~~
\hat{G}_{2}^{(3)} = \begin{pmatrix}
    1 & 0 & 0 \\
    0 & -1 & 0 \\
    0 & 0 & 0
  \end{pmatrix},~~\dots
\end{align}
and all of them have the energy spectrum being $\Lambda = \{ 1, 0, -1\}$. This correspondingly leads to spectral gaps that include $\widebar{\Gamma} = \{1,2\}$ unique positive distances. The recipe for taking the corresponding derivative for $\hat{U}(x) = \exp(-i x \hat{G}^{(3)}/2)$ as a part of function encoding (or energy) then relies on $S=2$ differentiation rules. The expression thus is the same as for $\theta$ derivative of the \texttt{fSim} gate,
\begin{align}
\label{eq:df_qutrit}
\frac{df(x)}{dx} &= \alpha_1(\delta_{1,2}, \Delta_{1,2})\Big[f(x+\delta_1) - f(x-\delta_1)\Big] + \\ \notag &~+ \alpha_2(\delta_{1,2}, \Delta_{1,2}) \Big[f(x+\delta_2) - f(x-\delta_2)\Big],
\end{align}
where $\alpha_{1,2}$ coefficients are the same as in Eqs.~\eqref{eq:alpha_1}--\eqref{eq:alpha_2} for gaps $\Delta_1 = 1$ and $\Delta_2 = 2$. As for the variance, the derivative for qutrit circuits measured with a limited number of shots follows Eq.~\eqref{eq:var_fSim}. It is minimized for $\{\delta_1, \delta_2\}$ with absolute values of $0.80 \pi$ and $0.29 \pi$.


\subsection{Hardware discussion}

Until recently quantum computing has been dominated mostly by the digital paradigm. From the theoretical perspective, this relies on the realization of a universal gate set, and specifically targeting its error-corrected operation. One example is a minimal Clifford + T gate set, which is universal --- the noise introduced by each gate can be suppressed exponentially using error correction protocols. However, this introduces a large overhead, mostly defined by the qubit rotation synthesis via Solovay-Kitaev approach. 

For current noisy intermediate scale quantum computers the successful operation usually relies on more expressive circuits. In this case the gates correspond to single-qubit rotations and entangling operations, implemented in hardware without synthesis. While more conventional entangling gates correspond to \texttt{CNOT}, \texttt{CZ}, and $\sqrt{\texttt{iSWAP}}$, hardware teams become increasingly enthusiastic about hardware-efficient gatesets that avoid overheads in implementation and have higher fidelity. In fact, this understanding has led to the demonstration of quantum supremacy, where the implementation of \texttt{fSim}-type gates and the \texttt{Sycamore} gate brought crucial advantage over \texttt{CZ}-based random circuits precisely because of their richer eigenspectrum and higher entangling power~\cite{Arute2019}. Another example is the cross-resonance \texttt{CR} gate that enabled cloud-based quantum computing~\cite{Kandala2017}.

It was realized that some architectures have difficulties in simulating effective `traditional' gatesets, while being more effective in evolving their system (or qubits) over some more complex Hamiltonian with a pulse-controlled evolution. Examples include global entangling gates native to trapped ion systems~\cite{Maslov2018}, analog blocks as realized in neutral atoms~\cite{Henriet2020,Scholl2021} as well as native three-qubit gates realized via Rydberg blockade~\cite{Levine2019}. The burgeoning development of native gate sets and its high performance led to the growth of the digital-analog quantum computing paradigm~\cite{Sanz2020,Sanz2021}.

At the same time, from an algorithmic design perspective, many near-term algorithms depend in one way or another on a variational quantum circuit ansatz, that is variationally optimized in a hybrid-classical feedback loop in order to optimize the system's output state for some condition or salient feature~\cite{Cerezo2021}. In some cases, there is a particular problem-guided structure to these ansatze, requiring particular Hamiltonians to be evolved for a fixed time (including QAOA \cite{Farhi2014}, variational phase estimation \cite{obrien2021}, variational fast forwarding \cite{cirstoiu2020} \textit{etc}). In other cases, one simply wishes to explore a large part of the Hilbert-space in an over-parametrization setting, and one mostly cares about short-depth entangling/expressive power \cite{Kandala2017, Mitarai2018, Sanz2020, Elfving2021, Kyriienko2021}. In all hardware-efficient strategies, this means the native gate sets can be chosen directly to avoid compilation overhead, and their differentiation is highly required. 

Whether driven by hardware or algorithmic considerations, it is clear there is a need in the art for differentiating gates other than those generated by single-Pauli-string generators. Therefore, we see the main advantage of generalizing parameter shift rules to generators with complex eigenspectrum as a valuable tool for variational protocols realized with hardware-native gates, positively contributing to the success of their operation.


\section{Conclusions} 

To conclude, we proposed an approach for differentiating quantum circuits where unitaries have generic generators. It can treat generators with multiple unique spectral gaps $S$, and generalizes the parameter shift rule, which is limited to generators with at most two unique eigenvalues ($S=1$). Using a spectral decomposition, we provide simple formulae for measuring the derivative as a sum of weighted function measurements. Considering symmetric shifts (pairs of equal magnitude and different signs), we provide a measurement schedule with at most $2S$ function evaluations, needed to learn $S$ complex matrix elements in a generic system. We also provide an approach with distinct (non-symmetric) shifts, where $2S+1$ measurements are required. The developed approach is applied to various quantum gates, including fSim gate ($S=2$), cross-resonance gate, and qutrit rotations. We provide explicit expressions for derivatives and analyze the variance for measuring derivative with a finite number of shots. Our work aims to enrich the differentiation strategies for hardware-efficient variational algorithms and quantum embeddings of derivatives.\vspace{1mm}

\textit{Note added.---}During the final stages of completing this work, several preprints were submitted on the same subject~\cite{Izmaylov2021,Wierichs2021}. In \cite{Izmaylov2021} the authors also considered a spectral decomposition, but concentrated on generator decomposition approaches that can offer beneficial measurement scaling for some cases. In \cite{Wierichs2021} the authors develop general rules that include the stochastic parameter-shift rule, and additionally concentrate on the case of spectra with equidistant eigenvalues.\\

\appendix

\section{Coefficients for $S=3$ differentiation rules}
\label{A}

Circuit differentiation rules for a case of the generator with three distinct positive gaps ($S=3$) require six measurements for symmetric shifts. The derivative is presented in the main text in Eq.~\eqref{eq:dfdx_CR}, and the coefficients are
\begin{align}
\label{eq:nu_1}
\nu_1 &= \Delta_3 \left[ \sin \left(\frac{\delta_{2}
   \Delta_2}{2}\right) \sin\left(\frac{\delta_{3} \Delta_1}{2}\right) - \sin \left(\frac{\delta_{2} \Delta_1}{2}\right) \sin \left(\frac{\delta_{3} \Delta_2}{2}\right) \right] \\ \notag & +\Delta_2 \left[ \sin \left(\frac{\delta_{2} \Delta_1}{2}\right)
   \sin \left(\frac{\delta_{3} \Delta_3}{2}\right)-
   \sin \left(\frac{\delta_{2} \Delta_3}{2}\right) \sin \left(\frac{\delta_{3} \Delta_1}{2}\right) \right] \\ \notag & + \Delta_1 \left[ \sin \left(\frac{\delta_{2} \Delta_3}{2}\right) \sin \left(\frac{\delta_{3} \Delta_2}{2}\right)- \sin \left(\frac{\delta_2 \Delta_2}{2}\right) \sin \left(\frac{\delta_{3} \Delta_3}{2}\right) \right],
\end{align}
\begin{align}
\label{eq:nu_2}
\nu_2 &= \Delta_{3} \left[ \sin
   \left(\frac{\delta_{1} \Delta_{1}}{2}\right) \sin \left(\frac{\delta_3 \Delta_{2}}{2}\right)- \sin \left(\frac{\delta_{1} \Delta_{2}}{2}\right) \sin \left(\frac{\delta_{3} \Delta_1}{2}\right) \right] \\ \notag &+\Delta_{2} \left[ \sin \left(\frac{\delta_{1} \Delta_3}{2}\right) \sin \left(\frac{\delta_{3} \Delta_{1}}{2}\right)- \sin \left(\frac{\delta_{1}
   \Delta_{1}}{2}\right) \sin \left(\frac{\delta_{3} \Delta_{3}}{2}\right) \right] \\ \notag &+\Delta_{1}
   \left[ \sin \left(\frac{\delta_{1} \Delta_{2}}{2}\right) \sin \left(\frac{\delta_{3} \Delta_3}{2}\right) - \sin \left(\frac{\delta_{1} \Delta_{3}}{2}\right) \sin
   \left(\frac{\delta_{3} \Delta_{2}}{2}\right) \right],
\end{align}
\begin{align}
\label{eq:nu_3}
\nu_3 &= \Delta_{3} \left[ \sin \left(\frac{\delta_{1} \Delta_{2}}{2}\right) \sin \left(\frac{\delta_{2} \Delta_{1}}{2}\right) - \sin \left(\frac{\delta_1 \Delta_{1}}{2}\right) \sin \left(\frac{\delta_{2} \Delta_{2}}{2}\right) \right] \\ \notag & +\Delta_{2} \left[ \sin \left(\frac{\delta_1 \Delta_{1}}{2}\right) \sin \left(\frac{\delta_{2} \Delta_{3}}{2}\right) - \sin \left(\frac{\delta_{1} \Delta_{3}}{2}\right) \sin \left(\frac{\delta_{2} \Delta_{1}}{2}\right) \right] \\ \notag & +\Delta_{1} \left[ \sin \left(\frac{\delta_{1} \Delta_{3}}{2}\right)
   \sin \left(\frac{\delta_{2} \Delta_{2}}{2}\right) -  \sin \left(\frac{\delta_{1} \Delta_{2}}{2}\right)
   \sin \left(\frac{\delta_{2} \Delta_{3}}{2}\right) \right]. 
\end{align}
The denominator reads
\begin{align}
\label{eq:Nu}
\mathcal{V} &= \sin \left(\frac{\delta_{1} \Delta_{3}}{2}\right) \sin \left(\frac{\delta_{2} \Delta_{2}}{2}\right) \sin \left(\frac{\delta_{3} \Delta_{1}}{2}\right) \\ \notag &-\sin
   \left(\frac{\delta_{1} \Delta_{3}}{2}\right) \sin \left(\frac{\delta_2 \Delta_{1}}{2}\right) \sin \left(\frac{\delta_{3} \Delta_{2}}{2}\right) \\ \notag &+\sin \left(\frac{\delta_{1} \Delta_{2}}{2}\right) \sin
   \left(\frac{\delta_{2} \Delta_{1}}{2}\right) \sin \left(\frac{\delta_{3} \Delta_3}{2}\right) \\ \notag &-\sin \left(\frac{\delta_{1}
   \Delta_{2}}{2}\right) \sin
   \left(\frac{\delta_{2} \Delta_{3}}{2}\right) \sin \left(\frac{\delta_{3} \Delta_{1}}{2}\right) \\ \notag &+\sin \left(\frac{\delta_{1} \Delta_1}{2}\right) \sin \left(\frac{\delta_{2} \Delta_{3}}{2}\right) \sin \left(\frac{\delta_{3} \Delta_{2}}{2}\right) \\ \notag &-\sin \left(\frac{\delta_{1} \Delta_{1}}{2}\right) \sin \left(\frac{\delta_{2}
   \Delta_{2}}{2}\right) \sin \left(\frac{\delta_{3} \Delta_{3}}{2}\right).
\end{align}
The variance for $S=3$ can be obtained analogously to the considered $S=1$ and $S=2$ cases. It corresponds to variances of function measurements, weighted by coefficients $\nu_s^2/\mathcal{V}^2$.



\begin{thebibliography}{99}

\bibitem{HHL2009} A. W. Harrow, A. Hassidim, and S. Lloyd, Quantum Algorithm for Linear Systems of Equations, Phys. Rev. Lett. \textbf{103}, 150502 (2009).

\bibitem{Childs2017} A. M. Childs, R. Kothari, and R. D. Somma, Quantum algorithm for systems of linear equations with exponentially improved dependence on precision, SIAM Journal on Computing \textbf{46}, 1920 (2017).

\bibitem{Lloyd2014} S. Lloyd, M. Mohseni, and P. Rebentrost, Quantum principal component analysis, Nature Physics \textbf{10}, 631 (2014).

\bibitem{Rebenfrost2014} P. Rebentrost, M. Mohseni, and S. Lloyd, Quantum Support Vector Machine for Big Data Classification, Phys. Rev. Lett. \textbf{113}, 130503 (2014).

\bibitem{SchuldRev} M. Schuld, I. Sinayskiy, and F. Petruccione, An introduction to quantum machine learning, Contemp. Phys. \textbf{56}, 172 (2015).

\bibitem{Biamonte2017} J. Biamonte, P. Wittek, N. Pancotti, P. Rebentrost, N. Wiebe, and S. Lloyd, Quantum machine learning, Nature \textbf{549}, 195 (2017).


\bibitem{Havlicek2019} V. Havli\v{c}ek, A. D. Corcoles, K. Temme, A. W. Harrow, A. Kandala, J. M. Chow, and J. M. Gambetta, Supervised learning with quantum-enhanced feature spaces, Nature (London) \textbf{567}, 209 (2019).

\bibitem{SchuldBocharov2020} M. Schuld, A. Bocharov, K. M. Svore, and N. Wiebe, Circuit-centric quantum classifiers, Phys. Rev. A \textbf{101}, 032308 (2020).

\bibitem{Du2021} Yuxuan Du, Min-Hsiu Hsieh, Tongliang Liu, Dacheng Tao, and Nana Liu, Quantum noise protects quantum classifiers against adversaries, Phys. Rev. Research \textbf{3}, 023153 (2021).

\bibitem{Xia2021} Yi Xia, Wei Li, Quntao Zhuang, and Zheshen Zhang, Quantum-Enhanced Data Classification with a Variational Entangled Sensor Network, Phys. Rev. X \textbf{11}, 021047 (2021).

\bibitem{Nghiem2021} Nhat A. Nghiem, Samuel Yen-Chi Chen, and Tzu-Chieh Wei, Unified framework for quantum classification, Phys. Rev. Research \textbf{3}, 033056 (2021).

\bibitem{Samuel2021} Samuel Yen-Chi Chen, Chih-Min Huang, Chia-Wei Hsing, Ying-Jer Kao, An end-to-end trainable hybrid classical-quantum classifier, Mach. Learn.: Sci. Technol. \textbf{2}, 045021 (2021).


\bibitem{JGLiu2018} Jin-Guo Liu and Lei Wang, Differentiable Learning of Quantum Circuit Born Machines, Phys. Rev. A \textbf{98}, 062324 (2018).

\bibitem{Zeng2019} Jinfeng Zeng, Yufeng Wu, Jin-Guo Liu, Lei Wang, and Jiangping Hu, Learning and inference on generative adversarial quantum circuits, Phys. Rev. A \textbf{99}, 052306 (2019).

\bibitem{Benedetti2019b} M. Benedetti, D. Garcia-Pintos, O. Perdomo, V. Leyton-Ortega, Y. Nam, and A. Perdomo-Ortiz, A generative modeling approach for benchmarking and training shallow quantum circuits, npj Quantum Inf. \textbf{5}, 45 (2019). 

\bibitem{Zoufal2019} C. Zoufal, A. Lucchi, and S. Woerner, Quantum Generative Adversarial Networks for learning and loading random distributions, npj Quantum Inf. \textbf{5}, 103 (2019).

\bibitem{Coyle2020} B. Coyle, D. Mills, V. Danos, and E. Kashefi, The Born supremacy: quantum advantage and training of an Ising Born machine, npj Quantum Inf. \textbf{6}, 60 (2020).

\bibitem{Romero2021} J. Romero and A. Aspuru-Guzik, Variational quantum generators: Generative adversarial quantum machine learning for continuous distributions, Adv. Quantum Technol. \textbf{4}, 2000003 (2021).

\bibitem{Paine2021} A. E. Paine, V. E. Elfving, and O. Kyriienko, Quantum Quantile Mechanics: Solving Stochastic Differential Equations for Generating Time-Series, arXiv:2108.03190 [quant-ph] (2021).

\bibitem{Huang2020} He-Liang Huang et al., Experimental Quantum Generative Adversarial Networks for Image Generation, Phys. Rev. Applied \textbf{16}, 024051 (2021).


\bibitem{DunjkoRev} V. Dunjko and H. J. Briegel, Machine learning \& artificial intelligence in the quantum domain: a review of recent progress, Rep. Prog. Phys. \textbf{81}, 074001 (2018).

\bibitem{Melnikov2018} A. A. Melnikov, H. Poulsen Nautrup, M. Krenn, V. Dunjko, M. Tiersch, A. Zeilinger, and H. J. Briegel, Active learning machine learns to create new quantum experiments, PNAS \textbf{115}, 1221 (2018).

\bibitem{Saggio2021} V. Saggio, B. E. Asenbeck, A. Hamann, T. Strömberg, P. Schiansky, V. Dunjko, N. Friis, N. C. Harris, M. Hochberg, D. Englund, S. W\"{o}lk, H. J. Briegel, and P. Walther, Experimental quantum speed-up in reinforcement learning agents, Nature \textbf{591}, 229 (2021).


\bibitem{Lubasch2020} M. Lubasch, Jaewoo Joo, P. Moinier, M. Kiffner, and D. Jaksch, Variational quantum algorithms for nonlinear problems, Phys. Rev. A 101, 010301(R) (2020).

\bibitem{Kyriienko2021} O. Kyriienko, A. E. Paine, and V. E. Elfving, Solving nonlinear differential equations with differentiable quantum circuits, Phys. Rev. A \textbf{103}, 052416 (2021).

\bibitem{Knudsen2020} M. Knudsen and C. B. Mendl, Solving Differential Equations via Continuous-Variable Quantum Computers, arXiv:2012.12220 [quant-ph]

\bibitem{Garcia-Molina2021} P. Garcia-Molina, J. Rodriguez-Mediavilla, and J. J. Garcia-Ripoll, Solving partial differential equations in quantum computers, arXiv:2104.02668 [quant-ph]


\bibitem{Benedetti2019rev} M. Benedetti, E. Lloyd, S. Sack, and M. Fiorentini, Parameterized quantum circuits as machine learning models, Quantum Sci. Technol. \textbf{4}, 043001 (2019).

\bibitem{CerezoRev} M. Cerezo, A. Arrasmith, R. Babbush, S. C. Benjamin, S. Endo, K. Fujii, J. R. McClean, K. Mitarai, Xiao Yuan, L. Cincio, and P. J. Coles, Variational Quantum Algorithms, Nat. Rev. Phys. \textbf{3}, 625 (2021).

\bibitem{Cerezo2021} M. Cerezo, A. Sone, T. Volkoff, L. Cincio, and P. J. Coles, Cost function dependent barren plateaus in shallow parametrized quantum circuits, Nature Commun. \textbf{12}, 1791 (2021).

\bibitem{BhartiRev} K. Bharti, A. Cervera-Lierta, Thi Ha Kyaw, T. Haug, S. Alperin-Lea, Abhinav Anand, M. Degroote, H. Heimonen, J. S. Kottmann, T. Menke, Wai-Keong Mok, Sukin Sim, Leong-Chuan Kwek, and A. Aspuru-Guzik, Noisy intermediate-scale quantum (NISQ) algorithms, arXiv:2101.08448 [quant-ph]

\bibitem{SchuldPRL} M. Schuld and N. Killoran, Quantum Machine Learning in Feature Hilbert Spaces, Phys. Rev. Lett. \textbf{122}, 040504 (2019).

\bibitem{Goto2020} Takahiro Goto, Quoc Hoan Tran, and Kohei Nakajima, Universal Approximation Property of Quantum Feature Map, Phys. Rev. Lett. 127, 090506 (2021).

\bibitem{Schuld2021} M. Schuld, R. Sweke, and J. J. Meyer, The effect of data encoding on the expressive power of variational quantum machine learning models, Phys. Rev. A \textbf{103}, 032430 (2021).

\bibitem{LeCun2015} Y. LeCun, Y. Bengio, G. Hinton, Deep learning, Nature \textbf{521}, 436 (2015).

\bibitem{OBrien2019} T. E. O’Brien, B. Senjean, R. Sagastizabal, X. Bonet-Monroig, A. Dutkiewicz, F. Buda, L. DiCarlo, and L. Visscher, Calculating energy derivatives for quantum chemistry on a quantum computer, npj Quantum Inf. \textbf{5}, 113 (2019).

\bibitem{Mitarai2020} K. Mitarai, Y. O. Nakagawa, and W. Mizukami, Theory of analytical energy derivatives for the variational quantum eigensolver, Phys. Rev. Res. \textbf{2}, 013129 (2020).


\bibitem{Harrow2019} A. Harrow and J. Napp, Low-depth gradient measurements can improve convergence in variational hybrid quantum-classical algorithms, Phys. Rev. Lett. \textbf{126}, 140502 (2021); orginally appeared as arXiv:1901.05374 [quant-ph] in (2019)

\bibitem{Ekert2002} A. K. Ekert, C. M. Alves, D. K. L. Oi, M. Horodecki, P. Horodecki, and L. C. Kwek, Direct Estimations of Linear and Nonlinear Functionals of a Quantum State, Phys. Rev. Lett. \textbf{88}, 217901 (2002).

\bibitem{Higgott2019} O. Higgott, Daochen Wang, S. Brierley, Variational Quantum Computation of Excited States, Quantum \textbf{3}, 156 (2019).

\bibitem{Mitarai2018} K. Mitarai, M. Negoro, M. Kitagawa, and K. Fujii, Quantum circuit learning, Phys. Rev. A \textbf{98}, 032309 (2018).

\bibitem{Schuld2019} M. Schuld, V. Bergholm, C. Gogolin, J. Izaac, and N. Killoran, Evaluating analytic gradients on quantum hardware, Phys. Rev. A \textbf{99}, 032331 (2019).

\bibitem{Li2017} Jun Li, Xiaodong Yang, Xinhua Peng, and Chang-Pu Sun, Hybrid Quantum-Classical Approach to Quantum Optimal Control, Phys. Rev. Lett. \textbf{118}, 150503 (2017).

\bibitem{Mitarai2019a} K. Mitarai and K. Fujii, Methodology for replacing indirect measurements with direct measurements, Phys. Rev. Res. \textbf{1}, 013006 (2019).

\bibitem{Kottmann2021} J. S. Kottmann, A. Anand, A. Aspuru-Guzik, A Feasible Approach for Automatically Differentiable Unitary Coupled-Cluster on Quantum Computers, Chem. Sci. \textbf{12}, 3497 (2021).

\bibitem{Mari2021b} A. Mari, T. R. Bromley, and N. Killoran, Estimating the gradient and higher-order derivatives on quantum hardware, Phys. Rev. A \textbf{103}, 012405 (2021).


\bibitem{Cerezo2021b} M. Cerezo and P. J. Coles, Higher order derivatives of quantum neural networks with barren plateaus, Quantum Sci. Technol. \textbf{6}, 035006 (2021).

\bibitem{Mari2021} A. Mari, T. R. Bromley, and N. Killoran, Estimating the gradient and higher-order derivatives on quantum hardware, Phys. Rev. A \textbf{103}, 012405 (2021).

\bibitem{Hubregtsen2021} T. Hubregtsen, F. Wilde, S. Qasim, and J. Eisert, Single-component gradient rules for variational quantum algorithms, arXiv:2106.01388 [quant-ph] (2021).

\bibitem{Vidal2018} J. G. Vidal and D. O. Theis, Calculus on parameterized quantum circuits, arXiv:1812.06323 [quant-ph] (2018).

\bibitem{Banchi2021} L. Banchi and G. E. Crooks, Measuring Analytic Gradients of General Quantum Evolution with the Stochastic Parameter Shift Rule, Quantum \textbf{5}, 386 (2021).

\bibitem{Crooks2019} G. E. Crooks, Gradients of parameterized quantum gates using the parameter-shift rule and gate decomposition, arXiv:1905.13311 [quant-ph]

\bibitem{Bespalova2021} T. A. Bespalova and O. Kyriienko, Hamiltonian operator approximation for energy measurement and ground state preparation, PRX Quantum \textbf{2}, 030318 (2021).

\bibitem{Kozin2019} V. K. Kozin and O. Kyriienko, Quantum Time Crystals from Hamiltonians with Long-Range Interactions, Phys. Rev. Lett. \textbf{123}, 210602 (2019).

\bibitem{Gard2019} B. T. Gard, L. Zhu, G. S. Barron, N. J. Mayhall, S. E. Economou, E. Barnes, Efficient symmetry-preserving state preparation circuits for the variational quantum eigensolver algorithm, npj Quantum Information \textbf{6}, 10 (2020).

\bibitem{Foxen2020} B. Foxen et al. (Google AI Quantum), Demonstrating a Continuous Set of Two-Qubit Gates for Near-Term Quantum Algorithms, Phys. Rev. Lett. \textbf{125}, 120504 (2020).


\bibitem{Mi2021} Xiao Mi, Pedram Roushan et al., Information Scrambling in Computationally Complex Quantum Circuits, arXiv:2101.08870 [quant-ph] (2021).

\bibitem{Anselmetti2021} G.-L. R. Anselmetti, D. Wierichs, C. Gogolin, and R. M. Parrish, Local, Expressive, Quantum-Number-Preserving VQE Ansatze for Fermionic Systems, arXiv:2104.05695 [quant-ph] (2021).


\bibitem{Chow2011} J. M. Chow, A. D. Corcoles, J. M. Gambetta, C. Rigetti, B. R. Johnson, J. A. Smolin, J. R. Rozen, G. A. Keefe, M. B. Rothwell, M. B. Ketchen, and M. Steffen, Simple All-Microwave Entangling Gate for Fixed-Frequency Superconducting Qubits, Phys. Rev. Lett. \textbf{107}, 080502 (2011).

\bibitem{Sheldon2016} S. Sheldon, E. Magesan, J. M. Chow, and J. M. Gambetta, Procedure for systematically tuning up cross-talk in the cross-resonance gate, Phys. Rev. A \textbf{93}, 060302(R) (2016).

\bibitem{Magesan2020} E. Magesan and J. M. Gambetta, Effective Hamiltonian models of the cross-resonance gate, Phys. Rev. A \textbf{101}, 052308 (2020).

\bibitem{Malekakhlagh2020} M. Malekakhlagh, E. Magesan, and D. C. McKay, First-principles analysis of cross-resonance gate operation, Phys. Rev. A \textbf{102}, 042605 (2020).

\bibitem{Kyriienko2018} O. Kyriienko, A. S. S{\o}rensen, Floquet quantum simulation with superconducting qubits, Phys. Rev. Appl. \textbf{9}, 064029 (2018).


\bibitem{Choi2020} Joonhee Choi, Hengyun Zhou, H. S. Knowles, R. Landig, Soonwon Choi, and M. D. Lukin, Robust Dynamic Hamiltonian Engineering of Many-Body Spin Systems, Phys. Rev. X \textbf{10}, 031002 (2020).

\bibitem{Arute2019} F. Arute et al., Quantum supremacy using a programmable superconducting processor, Nature (London) \textbf{574}, 505 (2019). 

\bibitem{Kandala2017} A. Kandala, A. Mezzacapo, K. Temme, M. Takita, M. Brink, J. M. Chow, and J. M. Gambetta, Hardware-efficient variational quantum eigensolver for small molecules and quantum magnets, Nature (London) \textbf{549}, 242 (2017).

\bibitem{Maslov2018} D. Maslov and Yunseong Nam, Use of global interactions in efficient quantum circuit constructions, New J. Phys. \textbf{20}, 033018 (2018).

\bibitem{Henriet2020} L. Henriet, L. Beguin, A. Signoles, T. Lahaye, A. Browaeys, G.-O. Reymond, and C. Jurczak, Quantum computing with neutral atoms, Quantum \textbf{4}, 327 (2020).

\bibitem{Scholl2021} P. Scholl, H. J. Williams, G. Bornet, F. Wallner, D. Barredo, T. Lahaye, A. Browaeys, L. Henriet, A. Signoles, C. Hainaut, T. Franz, S. Geier, A. Tebben, A. Salzinger, G. Z\"{u}rn, and M. Weidem\"{u}ller, Microwave-engineering of programmable XXZ Hamiltonians in arrays of Rydberg atoms, arXiv:2107.14459 [quant-ph] (2021).

\bibitem{Levine2019} H. Levine, A. Keesling, G. Semeghini, A. Omran, T. T. Wang, S. Ebadi, H. Bernien, M. Greiner, V. Vuletic, H. Pichler, and M. D. Lukin, Parallel Implementation of High-Fidelity Multiqubit Gates with Neutral Atoms, Phys. Rev. Lett. \textbf{123}, 170503 (2019).

\bibitem{Sanz2020} A. Parra-Rodriguez, P. Lougovski, L. Lamata, E. Solano, and M. Sanz, Digital-analog quantum computation, Phys. Rev. A \textbf{101}, 022305 (2020).

\bibitem{Sanz2021} P. Garcia-Molina, A. Martin, and M. Sanz, Noise in Digital and Digital-Analog Quantum Computation, arXiv:2107.12969 [quant-ph] (2021).

\bibitem{Farhi2014} E. Farhi, J. Goldstone, and S. Gutmann, A Quantum Approximate Optimization Algorithm, arXiv:1411.4028 [quant-ph] (2014).

\bibitem{obrien2021} T. E. O'Brien, S Polla, N C. Rubin, W. J. Huggins, S. McArdle, S. Boixo, J. R. McClean, and R. Babbush, Error mitigation via verified phase estimation, Phys. Rev. X \textbf{2}, 020317 (2021).

\bibitem{cirstoiu2020} C. C\^{i}rstoiu, Z. Holmes, J. Iosue, L. Cincio, P. J. Coles and A. Sornborger, Variational fast forwarding for quantum simulation beyond the coherence time, npj Quantum Inf \textbf{6}, 82 (2020).

\bibitem{Elfving2021} V. E. Elfving, M. Millaruelo, J. A. G\'amez and C. Gogolin, Simulating quantum chemistry in the seniority-zero space on qubit-based quantum computers, Phys. Rev. A \textbf{103}, 032605 (2021).


\bibitem{Izmaylov2021} A. F. Izmaylov, R. A. Lang, and Tzu-Ching Yen, Analytic gradients in variational quantum algorithms: Algebraic extensions of the parameter-shift rule to general unitary transformations, arXiv:2107.08131 [quant-ph] (2021).

\bibitem{Wierichs2021} D. Wierichs, J. Izaac, Cody Wang, and Cedric Yen-Yu Lin, General parameter-shift rules for quantum gradients, arXiv:2107.12390 [quant-ph] (2021).


\end{thebibliography}
\end{document}